\documentclass[twocolumn, showpacs, showkeys, aps]{revtex4-1}

\usepackage{amssymb,amsmath,xcolor,graphicx,xspace,colortbl,rotating} %
\usepackage[raggedrightboxes]{ragged2e} 
\usepackage{textcomp}
\usepackage{boxedminipage}  
\usepackage{float}  
\usepackage{graphics}  
\usepackage{ragged2e}  
\usepackage{tabulary}  
\usepackage{wrapfig}  
\usepackage{xcolor}  
\graphicspath{{TriangleARXIVa_graphics/}{TriangleARXIVa_tcache/}{TriangleARXIVa_gcache/}}
\DeclareGraphicsExtensions{.pdf,.eps,.ps,.png,.jpg,.jpeg}

\begin{document}
\title{Conductivity properties of the Sierpinski triangle
}
\author{Clinton DeW. Van Siclen}
\email{
cvansiclen@gmail.com
}
\affiliation{
1435 W 8750 N, Tetonia, Idaho 83452, USA}
\date{
\today
}
\begin{abstract}The classic Sierpinski triangle comprised of conducting bonds is multifractal. Thus the critical exponents and dimensions related to the conductivity are obtained asymptotically---that is, in the limit that the correlation length $\xi $ of the recursive triangle goes to infinity.
\end{abstract}
\maketitle

\section{
Introduction}

\begin{figure}[b]\centering 
\setlength\fboxrule{0.01in}\setlength\fboxsep{0.1in}\fcolorbox[HTML]{FFFFFF}{FFFFFF}{\includegraphics[ width=2in, height=1.7288135593220344in,]{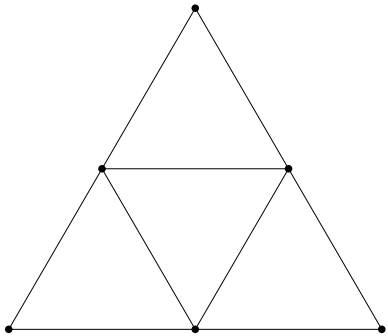}
}
\caption{Sierpinski triangle T\textsubscript {$1$}, which is the generator for triangle T\textsubscript {$i$}.
}\label{fig1}\end{figure}

\begin{figure}[b]\centering 
\setlength\fboxrule{0.01in}\setlength\fboxsep{0.1in}\fcolorbox[HTML]{FFFFFF}{FFFFFF}{\includegraphics[ width=2in, height=1.7272727272727273in,]{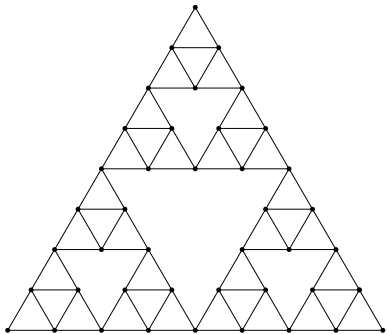}
}
\caption{Sierpinski triangle T\textsubscript {$3$}.
}\label{fig2}\end{figure}

The generator (iteration $i =1$) of the classic bond-and-node Sierpinski triangle (AKA gasket or sieve) is shown in Fig. \ref{fig1};
the triangle at iteration $i =3$ is shown in Fig. \ref{fig2}.
 Visually the triangle is self-similar but is actually multifractal: A triangle T\textsubscript {$i$} has side length $2^{i}$ (in unit of bond length) and contains $3^{i +1}$ bonds. Clearly there is no numerical value $d_{f}$ that satisfies the equation $3^{i +1} =(2^{i})^{d_{f}}$ for all values of $i$. 

The physical property of interest is the conductivity $\sigma ^{(i)}$ of the triangle T\textsubscript {$i$}. In this paper the conductivity of the recursive Sierpinski triangle is shown to be a critical phenomenon, and asymptotic values for the power-law exponents are derived or calculated.

The following section briefly describes the Walker Diffusion Method \cite{cvs99,cvs02} by which the analytical and numerical results are obtained. Then Section III applies the WDM to the bond-and-node Sierpinski triangle. With these results in hand, Section IV discusses results in the literature that, by and large, conflict fundamentally with those reported here. Lastly, Section V presents some final comments on this work.

\section{
Walker Diffusion Method}
The WDM was developed to calculate effective transport coefficients (e.g., conductivity) of composite materials and systems \cite{cvs99,cvs02}. This method exploits the isomorphism between the transport equations and the diffusion equation for a collection of non-interacting walkers (hence the name).

Accordingly, the walkers reside on the zero-dimensional nodes (or vertices) of a regular network of bonds (or edges) \cite{cvs02}. The principle of detailed balance ensures that at equilibrium (i.e., no walker sources or sinks) a uniform density $\rho _{i} =1$ of walkers is maintained. This is implemented by a ``variable residence time'' algorithm whereby every attempted move is successful but the move is accomplished over a variable time interval. Specifically, the direction of each move from a node $i$ (to an adjacent node $j$) is determined randomly by the set of probabilities $\{P_{ij}\}$, where

\begin{equation}P_{ij} =\frac{\sigma _{ij}}{\sum _{k}\sigma _{ik}} \label{e1}
\end{equation}
and the set $\{\sigma _{ik}\}$ are the conductivities of the bonds connecting node $i$ and adjacent nodes $k$. The time interval over which the move occurs is

\begin{equation}T_{i} =\frac{\phi }{\sum _{k}\sigma _{ik}} \label{e2}
\end{equation}where $\phi  =1$ in the case of orthogonal networks (e.g., square and cubic networks) and $\phi  =3/2$ in the case of triangular networks. The path of the walker thus reflects the distribution and conductivity of the conducting bonds, and may be described at the macroscopic scale by the diffusion coefficient $D_{w}$. That is related to the effective conductivity $\sigma $ by

\begin{equation}\sigma  =f_{w}\thinspace D_{w} \label{e3}
\end{equation}where the factor $f_{w}$ is the \textit{fraction} of walkers that are mobile (so equal to the fraction of nodes that have an attached conducting bond). The value $D_{w}$ is calculated from the equation

\begin{equation}D_{w} =\frac{\left \langle R(t)^{2}\right \rangle }{2dt} \label{e4}
\end{equation}where $d$ is the Euclidean dimension of the network; and the set $\{R\}$ of walker displacements, each occurring over the time interval $t$, comprises a Gaussian distribution that must necessarily be centered well beyond $\xi $. (For practical purposes, the correlation length $\xi $ is the length scale above which the ``effective'', or macroscopic, value of a transport property is obtained.)

For displacements $R <\xi $, the walker diffusion is anomalous rather than Gaussian due to the heterogeneity of the system at length scales less than $\xi $. There is, however, an additional characteristic length $\xi _{0} <\xi $ at which the system is effectively homogeneous; this may correspond, for example, to the size of a characteristic repeating unit in the system. A walker displacement of $\xi $ requiring a travel time $t_{\xi } =\xi ^{2}/(2dD_{w})$ is then comprised of $(\xi /\xi _{0})^{d_{w}}$ segments of length $\xi _{0}$, each requiring a travel time of $t_{0} =\xi _{0}^{2}/(2dD_{0})$, where $D_{0}$ is the walker diffusion coefficient calculated for the unit of size $\xi _{0}$. Setting $t_{\xi } =(\xi /\xi _{0})^{d_{w}}t_{0}$ gives the relation

\begin{equation}D_{w} =D_{0}\genfrac{(}{)}{}{}{\xi }{\xi _{0}}^{2 -d_{w}} \label{e5}
\end{equation}between the walker diffusion coefficient $D_{w}$, the fractal dimension $d_{w}$ of the walker path, and the correlation length $\xi $ \cite{cvs99a}.

\section{
Conductivity and exponent relations
}
The length $\xi _{0}$ is the bond length $\lambda $, which is also the side length of the triangle T\textsubscript {$0$}. Thus $D_{0} =D_{w}^{(0)}$ which is the diffusion coefficient for the walker on T\textsubscript {$0$}. The coefficient value is obtained by noting that $D_{w}^{(0)}$ is the diffusion coefficient for a walker on an \textit{infinite array} of T\textsubscript {$0$} triangles, which of course is simply the regular triangular network. Thus $D_{0} =\sigma _{0}$, where $\sigma _{0}$ is the conductivity of the bonds comprising the Sierpinski triangle.

More generally, the diffusion coefficient for walkers on the multifractal triangle T\textsubscript {$i$} is

\begin{equation}D_{w}^{(i)} =D_{w}^{(i -1)}\genfrac{(}{)}{}{}{\xi ^{(i)}}{\xi ^{(i -1)}}^{2 -d_{w}^{(i)}} =D_{w}^{(i -1)}\thinspace 2^{2 -d_{w}^{(i)}} \label{g1}
\end{equation}where the exponent $d_{w}^{(i)}$ is the fractal dimension of the walker path comprised of segments of length $\xi ^{(i -1)} =2^{i -1}\lambda $; that is, of the size of T\textsubscript {$i -1$} (note that visually, T\textsubscript {$i$} is three T\textsubscript {$i -1$} triangles in a stacked, triangular arrangement). Equation (\ref{g1}) leads to the relation

\begin{equation}D_{w}^{(i)} =\sigma _{0}\thinspace 2^{2 -d_{w}^{(1)}\thinspace }2^{2 -d_{w}^{(2)}} \cdots \thinspace 2^{2 -d_{w}^{(i)}} =\sigma _{0}(2^{i})^{2 -\mu _{w}^{(i)}} \label{g2}
\end{equation}where the exponent $\mu _{w}^{(i)}$, given by

\begin{equation}\mu _{w}^{(i)} =\frac{1}{i}\sum \limits _{k =1}^{k =i}d_{w}^{(k)} \label{g3}
\end{equation}accounts for the multifractal nature of the walker paths over T\textsubscript {$i$}.

\begin{figure}\centering 
\setlength\fboxrule{0.01in}\setlength\fboxsep{0.1in}\fcolorbox[HTML]{FFFFFF}{FFFFFF}{\includegraphics[ width=2in, height=1.7241379310344829in,]{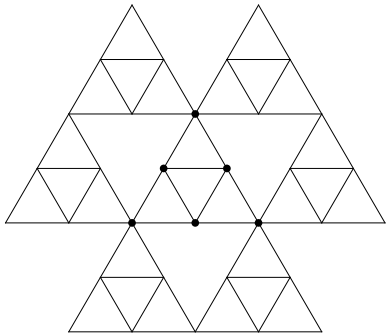}
}
\caption{Piece of an infinite 2D array of T\textsubscript {$1$} Sierpinski triangles. Non-conducting bonds are not shown.
}\label{fig3}\end{figure}

The fraction $f_{w}^{(i)}$ of nodes that connect the bonds comprising the T\textsubscript {$i$} triangle is obtained by considering an infinite array of T\textsubscript {$i$} triangles that fills 2D space and that has conductivity $\sigma ^{(i)}$. Such an array of T\textsubscript {$1$} triangles is shown in Fig. \ref{fig3}. Thus $f_{w}^{(i)}$ is given by the formula
\begin{equation}f_{w}^{(i)} =\frac{1 +3\sum \limits _{k =0}^{k =i -1}3^{k}}{4^{i}} =\frac{3^{i +1} -1}{2^{2i +1}} \label{h1}
\end{equation}(note that the denominator $4^{i}$ is the total number of network nodes in the periodic rhombus-shaped area that contains one T\textsubscript {$i$} triangle). Note that $f_{w}^{(i)} \approx \frac{3}{2}(\frac{3}{4})^{i} \sim (\frac{3}{4})^{i} =(2^{i})^{ -\gamma }$ where the exponent $\gamma  =\frac{\ln (4/3)}{\ln 2} =2 -\frac{\ln 3}{\ln 2}$.

Combining Eqs. (\ref{e3}) and (\ref{h1}) and (\ref{g2}) shows the conductivity of triangle T\textsubscript {$i$} to be

\begin{equation}\sigma ^{(i)} =\sigma _{0}\thinspace f_{w}^{(i)}\thinspace (2^{i})^{2 -\mu _{w}^{(i)}} . \label{e6}
\end{equation}Thus the conductivity obeys the asymptotic relation

\begin{equation}\sigma (\xi ) \sim \xi ^{ -\gamma }\thinspace \xi ^{2 -\mu _{w}^{ \ast }} =\xi ^{ -(\gamma  -2 +\mu _{w}^{ \ast })} \label{e7}
\end{equation}verifying that the conductivity of the recursive Sierpinski triangle is a critical phenomenon. The exponent $\mu _{w}^{ \ast }$ is the limit of $\mu _{w}^{(i)}$ as iteration $i \rightarrow \infty $.

Exponent relations are found by recognizing that the fraction $p$ of bonds that comprise the Sierpinski triangle declines with iteration $i$ as

\begin{equation}p^{(i)} =\frac{3^{i +1}}{3 \cdot 4^{i}} =\genfrac{(}{)}{}{}{3}{4}^{i} =(2^{i})^{ -1/\nu } =(\xi ^{(i)})^{ -1/\nu } \label{e8}
\end{equation}where the critical exponent $\nu  =\frac{\ln 2}{\ln (4/3)} =(2 -\frac{\ln 3}{\ln 2})^{ -1}$. The conductivity then declines asymptotically as

\begin{equation}\sigma  \sim p^{t} \label{e10}
\end{equation}where the conductivity exponent $t =\nu (\gamma  -2 +\mu _{w}^{ \ast })$. Note that $\nu  =\gamma ^{ -1}$ so $t =1 +\nu (\mu _{w}^{ \ast } -2)$ and $\mu _{w}^{ \ast } =\frac{\ln 3}{\ln 2} +\frac{t}{\nu }$.

Numerical values for the exponents $\mu _{w}^{(i)}$ can be obtained via the relation $D_{w}^{(i)} =\sigma _{0}(2^{i})^{2 -\mu _{w}^{(i)}}$. This is rewritten (after setting $\sigma _{0} =1$ for convenience)

\begin{equation}\mu _{w}^{(i)} =2 -\frac{\ln D_{w}^{(i)}}{\ln (2^{i})} . \label{e11}
\end{equation}

Similarly, numerical values for the path dimensions $d_{w}^{(i)}$ can be obtained via Eq. (\ref{g1}). That is rewritten

\begin{equation}d_{w}^{(i)} =2 -\frac{\ln (D_{w}^{(i)}/D_{w}^{(i -1)})}{\ln 2} . \label{g4}
\end{equation}The set $\{d_{w}^{(i)}\}$ of walker path dimensions is the manifestation of the multifractal nature of the bond-and-node Sierpinski triangle.

The diffusion coefficient $D_{w}^{(i)}$ is calculated by Eq. (\ref{e4}), from walks of time $t \gg t_{\xi }^{(i)}$. The walkers diffuse over an \textit{infinite array} of T\textsubscript {$i$} triangles, constructed such that it has the same conductivity as the single T\textsubscript {$i$} triangle. An example array (that for triangle T\textsubscript {$1$}) is shown in Fig. \ref{fig3}.
 Note that it demonstrates the condition $p^{(i)} =(3/4)^{i}$.

In practice, the walk over the infinite array is accomplished with a single T\textsubscript {$i$} triangle. A walker that reaches one of the three apex nodes (the three vertices opposite the three sides of length $2^{i}\thinspace \xi _{0}$) is confronted with possible moves in six different directions (as apparent in Fig. \ref{fig3}). By recording only the \textit{direction} of each move in a walk, a walker that \textit{arrives} at an apex node may \textit{leave} from any one of the three apex nodes, dependent on the \textit{direction} of that move. Then the displacement $R$ is the vector sum of all those directed moves.

\begin{figure}[b]\centering 
\setlength\fboxrule{0.01in}\setlength\fboxsep{0.1in}\fcolorbox[HTML]{FFFFFF}{FFFFFF}{\includegraphics[ width=2.7in, height=2.0025in,]{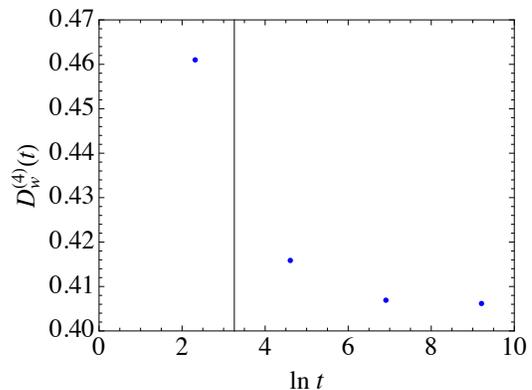}
}
\caption{$D_{w}^{(4)}(t)$ values calculated from walks of time $t$ over an infinite array of T\textsubscript {$4$} Sierpinski triangles.
}\label{fig4}\end{figure}

Figure \ref{fig4} shows the decline in $D_{w}^{(4)}$ value with increasing walk time $t$.
 The vertical line indicates the correlation time $t_{\xi }^{(4)}$ above which $D_{w}^{(4)}$ attains its true, macroscopic, value. [Note that $D_{w}^{(4)}$ continues to decline beyond $t_{\xi }^{(4)}$, as the tail of the probability distribution $P[R(t)]$ remains in the anomalous regime $R <\xi ^{(4)}$.]

Each point in Fig. \ref{fig4} (and corresponding points for triangles T\textsubscript {$1$}, T\textsubscript {$2$}, and T\textsubscript {$3$}) is obtained from $10$ sequences of $10^{6}$ walks of time $t$. [A \textit{sequence} of $10^{6}$ walks is actually a single, uninterrupted walk of time $10^{6} \times t$. During that long walk every displacement $R(t)$ is recorded, for a total of $10^{6}$ displacements.] The indicated value $D_{w}^{(4)}(t) =\left \langle R(t)^{2}\right \rangle /2dt$, where $\left \langle R(t)^{2}\right \rangle $ is the average of \textit{all} walks of time $t$ (that is, the average of all sequences). In every case the number of sequences is sufficient that an additional sequence would change the average value $\left \langle R(t)^{2}\right \rangle $ by only an insignificant amount (i.e., not affecting the point size in the figure).
\begin{table}\centering \caption{
Calculated values for the diffusion coefficient $D_{w}^{(i)}$ (with $\sigma _{0} =1$), the fractal path dimension $d_{w}^{(i)}$, and the multifractal-path exponent $\mu _{w}^{(i)}$ obtained from walks over the infinite 2D array of T\textsubscript {$i$} triangles.
}
\setlength\fboxrule{0pt}\setlength\fboxsep{0pt}\fcolorbox[HTML]{FFFFFF}{FFFFFF}{
\begin{tabular}[c]{p{13mm}p{31mm}p{23mm}p{13mm}}\hline
\hline
$i$ & \par $D_{w}^{(i)}$
& \par $d_{w}^{(i)}$
& $\mu _{w}^{(i)}$ \\
\hline
0 & 1 & 2 \\
1 & 0.650368(576) & 2.62067 & 2.62067 \\
2 & 0.526181(447) & 2.3057 & 2.46319 \\
3 & 0.455556(490) & 2.20793 & 2.3781 \\
4 & 0.406172(362) & 2.16554 & 2.32496 \\
\hline
\hline
\end{tabular}
}
\end{table}
\begin{table}\centering 
\setlength\fboxrule{0pt}\setlength\fboxsep{0pt}\fcolorbox[HTML]{FFFFFF}{FFFFFF}{
}
\end{table}\qquad 

Table I gives the calculated macroscopic $D_{w}^{(i)}$ values for triangle iterations $i =0 ,1 ,2 ,3 ,4$, and the corresponding $d_{w}^{(i)}$ and $\mu _{w}^{(i)}$ values calculated according to Eqs. (\ref{g4}) and (\ref{e11}), respectively. Clearly, the asymptotic value $\mu _{w}^{ \ast }$ for the multifractal-path exponent cannot be larger than $d_{w}^{(4)}$. A very crude extrapolation from the values $d_{w}^{(2)}$, $d_{w}^{(3)}$, and $d_{w}^{(4)}$, performed by the Shanks transformation method (see Appendix A), produces $d_{w}^{(\infty )} =2.133$, and consequently $\mu _{w}^{(i)} \rightarrow 2.133$. Of course this limit value must be regarded as heuristic.

\section{
Discussion}
The analysis above is very different from the standard approach presented in the literature \cite{gefen1981,guyer1984,gefen1984,tait1988,schul2000,havlin2002}. The latter uses the ``triangle-star'' ($\mathrm{\Delta } \rightarrow $Y) transformation to obtain the effective resistance $R^{(i)}$ of triangle T\textsubscript {$i$} when a potential drop is imposed across two of the three apex vertices. It is found that the ratio $R^{(i +1)}/R^{(i)} =5/3$, which gives the general relation

\begin{equation}R^{(i)} =(2^{i})^{\zeta }\thinspace R^{(0)} \label{e12}
\end{equation}with exponent $\zeta  =\ln (5/3)/\ln 2 =0.736966$.

To compare the triangle-star approach to the WDM approach, it is recognized that the former actually generates an ``effective triangle T\textsubscript {$0$}''---call it T\textsubscript {$0$}\textsuperscript {$(i)$}---with bonds of length $\lambda $ and resistance $\genfrac{(}{)}{}{}{5}{3}^{i}\frac{\lambda }{\sigma _{0}}$. Thus triangle T\textsubscript {$0$}\textsuperscript {$(i)$} has conductivity $\genfrac{(}{)}{}{}{3}{5}^{i}\sigma _{0}$, and consequently so does triangle T\textsubscript {$i$}.

However, this result follows from the consideration of a potential drop across two of the apex vertices. Thus the conductivities derived from the triangle-star transformation differ from those obtained by the WDM (for example, $\sigma ^{(i)} =f_{w}^{(i)}\thinspace D_{w}^{(i)}$ with $D_{w}^{(i)}$ values taken from Table I) and are not correct. And clearly $\zeta $ cannot be the ``resistance exponent'' $\zeta _{R}$ \cite{stauffer} appearing in the asymptotic relation $\sigma (\xi ) \sim \xi ^{ -\zeta _{R}}$.

Further, aside from the problematic issue of the exponent $\zeta $, the standard approach \textit{assumes} the exponent relation $d_{w} =d_{f} +\zeta _{R}$ and so obtains the analytical value $d_{w} =\ln 5/\ln 2 =2.322$. But the bond-and-node Sierpinski triangle is multifractal (\textit{not} fractal with dimension $d_{f} =\frac{\ln 3}{\ln 2}$), and in any event Ohm's law must be obeyed which leads to Eq. (\ref{e3}) and, in the case of a fractal, the exponent relation $\frac{t}{\nu } =\gamma  -2 +d_{w}$. If $\gamma  =\nu ^{ -1} =2 -d_{f}$ then this relation becomes $\frac{t}{\nu } =d_{w} -d_{f}$.

[Note that the exponent relation $d_{w} =d_{f} +\frac{t}{\nu }$ is characteristic of recursive fractals embedded in 2D space and comprised of \textit{sites}. For example, it is \textit{derived} for the Sierpinski carpet in Ref. \cite{cvs2017}.]

The calculated values for the fractal path dimension $d_{w}$ reported in the literature may be compared with the values for the multifractal-path exponent $\mu _{w}$ shown in Table I. The former are actually quite close to the analytical value derived from the standard approach, and to the value $\mu _{w}^{(4)}$ in Table I. Given and Mandelbrot \cite{given1983} obtain $d_{w} =2.32019$ by Monte Carlo simulations. Each walk continued until a prescribed number of steps was reached, or until the walk reached an apex node. Guyer \cite{guyer1984} obtains $d_{w} =2.318$ from a length-scale renormalization procedure for handling the diffusion equation. Havlin and Ben-Avraham \cite{havlin2002} obtain $d_{w} =2.32(1)$ by an exact enumeration method that generates all possible random walks starting from a specified node. All walks were less than $250$ steps (according to their Fig. 5) and were confined to a single triangle. Lara \textit{et al.} \cite{lara2013} obtain $d_{w} =2.27$ from discrete-time quantum walks confined to a T\textsubscript {$8$} triangle.

These calculations of $d_{w}$ suffer from their reliance on Sierpinski triangles of low orders of recursion (iteration $i \ll \infty $). The same complaint applies to the WDM calculations reported here. Unfortunately the computational demands quickly become overwhelming: With each iteration the correlation length $\xi ^{(i)}$ increases by a factor of $2$, so that the required walk time $t \gg t_{\xi }$ increases by more than a factor of $4$. And the larger the walk time $t$, the broader the probability distribution $P[R(t)]$, so that many more walks must be taken to obtain an accurate value $\left \langle R(t)^{2}\right \rangle $.

\section{
Concluding remarks}
This work relies on (1) the recognition that the bond-and-node Sierpinski triangle T\textsubscript {$i$} is multifractal, and (2) the use of the conductivity relation $\sigma ^{(i)} =f_{w}^{(i)}D_{w}^{(i)}$ and the consequent exponent relation $ -\frac{t}{\nu } = -\gamma  +2 -\mu _{w}^{ \ast }$. The critical exponents and dimensions are obtained analytically or numerically by the Walker Diffusion Method. Comparison is made to work reported in the literature.

Note, finally, that a check on the method and the code written for this work is provided by a simple problem related to that of the Sierpinski triangle. Appendix B considers the conductivity of a bond-and-node equilateral triangle of side length $2^{i}\lambda $ (\textit{no} additional conducting bonds in this case), where an analytical solution is available.

\begin{acknowledgments}
I thank my colleague, Dr. Thomas Wood (Center for Advanced Energy Studies, Idaho Falls, Idaho), for arranging my access to the resources of the University of Idaho Library (Moscow, Idaho).\end{acknowledgments}

\appendix

\section{
Shanks transformation method}
An estimate of the limit $d_{\infty }$ of a sequence of values $d_{1} >d_{2} >d_{3} >0$ can be obtained from solution of the three equations

\begin{align}d_{1} =d_{\infty } +\alpha  \\
d_{2} =d_{\infty } +\alpha \beta  \\
d_{3} =d_{\infty } +\alpha \beta ^{2}\end{align}where $\alpha $ and $\beta $ are positive constants, and $\beta  <1$. The three equations with three unknowns produce the value

\begin{equation}d_{\infty } =\frac{d_{3}d_{1} -d_{2}d_{2}}{d_{3} +d_{1} -2d_{2}} .
\end{equation}

\section{Conductivity of an equilateral triangle
}
The equilateral triangle T\textsubscript {$i$} has side length $2^{i}\lambda $: each side is comprised of $2^{i}$ bonds of length $\lambda $ and conductivity $\sigma _{0}$. Logically, the conductivity $\sigma ^{(i)}$ of an infinite 2D array of T\textsubscript {$i$} triangles equals the conductivity of an infinite array of T\textsubscript {$0$} triangles comprised of bonds of length $\lambda $ and conductivity $\sigma _{0}/2^{i}$; that is, $\sigma ^{(i)} =\sigma _{0}/2^{i}$.

The fraction $f_{w}^{(i)}$ of walkers that diffuse over T\textsubscript {$i$} (i.e., the fraction of nodes that connect conducting bonds) is given by the formula

\begin{equation}f_{w}^{(i)} =\frac{1 +3(2^{i} -1)}{4^{i}} .
\end{equation}Thus a WDM calculation of the walker diffusion coefficient $D_{w}^{(i)}$ will obtain the analytical value $D_{w}^{(i)} =\sigma _{0}\thinspace 2^{ -i}\thinspace (f_{w}^{(i)})^{ -1}$.

\end{document}